\documentclass[a4paper,12pt]{article}
\usepackage{citesort}
\usepackage{ae}
\usepackage[square]{natbib}
\usepackage{graphicx}
\usepackage{amssymb}

\title{Spillover on supported catalysts -- Dynamic Monte Carlo study}
\author{Lukasz Cwiklik, Barbara Jagoda-Cwiklik, Marek Frankowicz
\\
\\Department of Theoretical Chemistry, Faculty of Chemistry
\\Jagiellonian University
\\Krakow, Poland
}

\begin{document}

\maketitle

\begin{abstract}
The kinetics of a $2A + B_2 = 2AB$ reaction on supported metal catalyst with spillover effects is investigated using Dynamic Monte Carlo simulations. In the presented model $A$ particles can adsorb reversibly on both metal clusters and the support whereas $B_2$ particles can adsorb dissociatively only on metal clusters. Particles $A$ can diffuse from the support to metal clusters and vice versa. The model describes CO oxidation on Pd/Al$_2$O$_3$ catalyst. Steady-state reaction rates and surface coverages are investigated as functions of partial pressures of reactants in the gas phase. Simulations show that spillover effects change the reaction rate dramatically by increasing it and widening the pressure window of the reaction. In systems with spillover the reaction driven by an adsorption channel, in addition to a spillover channel, is observed. The character of the dependence of the reaction rate on reactants' partial pressures can be controlled by changes of spillover and adsorption channel contributions.
\end{abstract}

\section{Introduction} \label{sec:intro}

Spillover is an effect accompanying heterogeneous surface reactions when inactive surface regions significantly influence the kinetics of the overall catalytic process. It is caused by the fact that those parts of the surface which are inactive in the surface reaction can be active for other processes occurring during the catalytic process, i.e. adsorption-desorption processes, and increase or decrease concentrations of either substrate or product particles on active parts of the surface through the surface diffusion. The spillover effect is observed and plays a significant role in catalytic reactions on supported metal catalysts \cite{Conner1995}.

The description of process dynamics in systems with spillover is complicated for several reasons. Firstly -- surface heterogeneity. In most cases the heterogeneity, because of the occurrence of local effects, makes it impossible to use methods based on global chemical kinetics. Secondly -- surface heterogeneity means not only differences in energetic properties of different surface regions but also involves geometrical heterogeneities. There are active centers with different energetic properties on the surface but additionally those centers are often arranged in different ways (for example active centers can form various clusters placed on the surface in various ways or active centers can be spread randomly all over the surface). The chemical and physical properties of the surface depend also on the arrangement of active centers. Thirdly -- sizes of surface structures and length scales characteristic for spillover processes are tenths or hundredths of nanometers.  Thus it is very complicated to use quantum chemical or molecular dynamics calculations to investigate such large systems and processes occurring in such long time scales. To investigate surface processes particularly in case of inhomogeneous surfaces and non-equilibrium conditions one can use Monte Carlo methods \cite{Broadbelt2000, Zhdanov2000} (Kinetic Monte Carlo or Dynamic Monte Carlo). Such simulation methods make it possible to investigate systems much larger then in the case of quantum chemical calculations and analyze processes with characteristic times much longer than in molecular dynamics simulations. These methods also let to include the influence of local effects occuring on the surface.

The current state of theoretical research of reactions with spillover effects includes both computer simulations and attempts at the analytical description of such processes' kinetics. In one of the most recent theoretical publications in this subject Zhdanov and Kasemo \cite{Zhdanov1997} derived analytical formulas to describe dynamics of a model reaction with spillover effect -- oxidation of CO on a supported metal catalyst (for example Pd/Al$_2$O$_3$). Oxidation reactions of this type are good examples of processes with spillover from a practical point of view -- it is relatively simple to investigate them experimentally and they are important as actual catalytic processes. However, an analytical description of considered process requires many assumptions.

Libuda, Hoffmann et al. \cite{Libuda2001, Hoffmann2001} presented experimental results and computer simulations of kinetics of CO oxidation on Pd/Al$_2$O$_3$ but in those papers only mean-field simulation methods have been used and because of that methodology spatial heterogeneities have been neglected.

Recently, Kovaliov \cite{Kovaliov2003} et al. investigated $2A + B_2 \rightarrow 2AB$ reaction with spillover using Monte Carlo simulations. In their work A adsorption was assumed as irreversible and they studied the influence of metal particle's morphology on the dynamics of the process. Our results are comparable with the ones obtained in their paper but in our work the metal particle's morphology is neglected because we wanted to focus on the spillover and to investigate its influence on the catalytic process.

In this paper non-equilibrium Dynamic Monte Carlo simulations were used to investigate the CO oxidation reaction with the spillover effect on a supported metal catalyst. This allowed us to analyze the heterogeneous process kinetics for a wide range of its parameters and to include spatial heterogeneities. Using our simulation techniques new types of behavior occurring in this system have been found.

\section{System description and methodology} \label{sec:sys_desc}

In this paper we considered the dynamics of the catalytic process that can be presented by the following set of chemical equations:
\begin{equation} \label{eq:a_inact_ads}
A_{(g)} + \circ \leftrightarrows A_{(ads)}^{\circ}
\end{equation}
\begin{equation} \label{eq:a_act_ads}
A_{(g)} + * \leftrightarrows A_{(ads)}
\end{equation}
\begin{equation} \label{eq:a_diff}
A_{(ads)}^{\circ} + * \leftrightarrows A_{(ads)} + \circ
\end{equation}
\begin{equation} \label{eq:b_ads}
B_{2 (g)} + ** \rightarrow 2B_{(ads)}
\end{equation}
\begin{equation} \label{eq:react}
A_{(ads)} + B_{(ads)} \rightarrow AB_{(g)} + **
\end{equation}
where: $A_{(g)}$,  $A_{(ads)}$, $A_{(ads)}^{\circ}$ -- substrate A particles in the gas phase, adsorbed on the active metal particle and adsorbed on the inactive support respectively; $*$, $\circ$ -- adsorption sites respectively on the active metal particle and on the support; $**$ -- two neighboring adsorption sites on the metal particle; $B_{2 (g)}$ -- substrate B particle in the gas phase; $B_{(ads)}$ -- B particle adsorbed on the metal particle after the dissociative adsorption of $B_2$ particle; $AB_{(g)}$ -- surface reaction product particle desorbed to the gas phase.

The catalytic process takes place on the surface of the supported metal catalyst. This surface consists of two different regions: the support inactive in the surface reaction (for example, corresponding to Al$_2$O$_3$) and the active particles (for example, corresponding to nanometer Pd crystallites) -- the only parts of the surface where the surface reaction given by Eq. \ref{eq:react} takes place. It was assumed that an A particle can adsorb, desorb and diffuse all over the surface. B$_2$ particles can adsorb exclusively on the active region where they dissociate and the product of the dissociation (B particles) stays on the active region. B particles can neither desorb nor diffuse to the inactive support. As mentioned above, this model can describe, for example, a CO oxidation reaction on a supported metal catalyst Pd/Al$_2$O$_3$ \cite{Zhdanov1997}.

If we neglect adsorption and diffusion (Eq. \ref{eq:a_inact_ads}. and \ref{eq:a_diff}.) we would get a well known ZGB model of CO oxidation reaction on a single facet of metal catalyst \cite{Ziff1986}. Three additional differences between these two models exist. In our case A adsorption is reversible, diffusion of A particles on the metal particle is present and the surface reaction is not immediate. In the ZGB model A adsorption was an irreversible process, surface diffusion was forbidden and the reaction occurred immediate after adsorption of A particle on the site adjacent to an adsorbed B particle. We will compare our results obtained for the system without support (i.e. without spillover) with the ZGB model in Sec. \ref{subsec:react_on_particle}.

Analytical expressions describing the dynamics of the process in question can be derived assuming high surface reaction rates and high dominance of one substrate in the gas phase \cite{Zhdanov1997}. The latter assumption means that the value of the partial pressure of A substrate in the gas phase ($p_A$) is near 0 or 1 (partial pressures of substrates obey $p_A + p_{B_2} = 1$). From analytical results one can conclude the process achieves a maximum rate in the medium range of pressures for various sets of system parameters. The maximum rate region is the most important from a practical point of view. However, analytical expressions derived for extreme values of reactants' partial pressures do not allow exact description of the medium pressures range. Our Dynamic Monte Carlo simulations allowed us to analyze the system behavior in the whole range of pressure values.

In this paper the case of well-separated active particles was considered, i.e. diffusion spheres of active particles did not overlap. The investigation of process dynamics in such system can be reduced to simulations of the system consisting of one active particle placed on a relatively large support region. In our simulations the catalyst surface was represented by a two-dimensional square lattice with 150$\times$150 nodes. In the middle of the lattice a smaller square consisting of 50$\times$50 nodes was distinguished and this square corresponded to the metal particle active in the surface reaction. The rest of the lattice nodes were assumed to be a support inactive in the surface reaction. Periodic boundary conditions were used to exclude the influence of the edges of simulation lattice.

In simulations it was assumed that above the surface there is a mixture of gaseous substrates A and B$_2$ with known partial pressures of both compounds $p_A$ and $p_{B_2}$. According to the reaction mechanism we assumed that particles of substrate A can adsorb with given probabilities on active and inactive lattice sites ($P_{A,ads,act}$ and $P_{A,ads,inact}$ respectively). Particles A also can desorb from each site with probabilities $P_{A,des,act}$ or $P_{A,des,inact}$ because, according to Eq. \ref{eq:a_act_ads} and \ref{eq:a_inact_ads}, adsorption of substrate A is reversible. To mimic surface diffusion, adsorbed A particles could move on the lattice. This was done through jumps of A particle between neighboring lattice sites with given probabilities. The four nearest adjacent sites to the considered one were regarded as neighboring (von Neumann neighborhood). The presence of two types of lattice sites (active -- \textit{act} and inactive -- \textit{inact}) allowed for four different types of jumps which could be characterized by different values of jump probabilities (from active to active site, active to inactive, inactive to inactive and inactive to active).

It was assumed that B$_2$ particles can adsorb only on active lattice sites and that adsorption is dissociative. Such an adsorption mechanism allows the adsorption of B$_2$ particles to take place only on those unoccupied active lattice sites with at least one unoccupied neighboring site -- B$_2$ adsorption requires the presence of two unoccupied neighboring active lattice sites. Furthermore it was assumed that, due to the dissociation,  B$_2$ adsorption is irreversible because in the presented model the probability of B$_2$ desorption is smaller than the probabilities of other processes (B$_2$ desorption requires, in addition to crossing the surface bonding barrier, additionally the meeting of two adjacent adsorbed B particles and also the crossing of an association process energy barrier). Adsorbed B particles occupied only active sites because they could not diffuse from the active region to the support. In the presented model only A particles could diffuse from the support to the active region and vice versa -- this diffusion is \textit{the spillover}. In simulations, the diffusion of B particles adsorbed on the active region was neglected because B particles cannot cross the active -- inactive region border and just that process (i.e. spillover) governs the overall catalytic process dynamics. Diffusion processes within the active region do not directly influence this dynamics.

If two adjacent active sites were occupied by A and B particles, irreversible surface reaction (Eq. \ref{eq:react}) could take place, an AB particle was formed and it was immediately removed from the lattice (corresponding to the immediate desorption of the product to the gas phase). The presence of neighboring A and B particles on the active region could be caused either by direct adsorption of A and B particles on active sites or by surface diffusion of A particles from inactive to active sites. Therefore A particles could be transported to active regions by two different mechanisms: by direct adsorption (adsorption channel) and by surface diffusion from the inactive region (spillover channel).

\section{Simulation algorithm} \label{sec:sim_alg}

At the beginning of each simulation, values of partial pressures of A and B$_2$ substrates and values of processes probabilities were assumed. Each simulation consisted of repeated identical steps. Each step started with the random choice of one site on the simulation lattice.
\begin{itemize}
	\item If the chosen lattice site was unoccupied (i.e. there was no particle adsorbed on it) an adsorption attempt was realized; either an A or B$_2$ particle was chosen with a probability equal to partial pressures of substrates in the gas phase.
	\begin{itemize}
		\item If an A particle was chosen, an adsorption attempt was realized with the probability $P_{A,ads,act}$ or $P_{A,ads,inact}$ (corresponding to the kind of the chosen lattice site). After this attempt, and independently of its results, the simulation step ended.
		\item If a B$_2$ particle was chosen, one additional neighboring site of the current one was randomly chosen because the B$_2$ adsorption required two unoccupied neighboring sites. If the chosen neighboring site was occupied, the simulation step ended. If it was empty, an adsorption attempt on both the current and the neighboring site was realized with the probability $P_{B,ads,act}$ or  $P_{B,ads,inact}$(in the presented model $P_{B,ads,inact} = 0$ because B$_2$ could adsorb only on the active region of the surface). After the B$_2$ adsorption attempt, and independently of its results, the simulation step ended.
	\end{itemize}
	\item If the lattice site chosen at the beginning of the simulation step was occupied and it contained an A particle, one of the three processes was chosen randomly: the diffusion, desorption or reaction.
	\begin{itemize}
		\item In the case of diffusion, one of four neighboring sites was randomly chosen and if it was occupied, the simulation step ended. If the neighboring site was empty, an attempt of the particle jump to this empty site was made with the probability depending on the kind of the starting and the target site. After this attempt simulation step ended.
		\item In the case of the desorption an attempt of removing the particle out of the lattice with the probability $P_{A,des,act}$ or $P_{A,des,inact}$ (depending on the kind of the lattice site) was made. After this attempt and independently of its results, the simulation step ended.
		\item In the case of reaction, the type of the current site was checked. If the current site was inactive one -- the simulation step ended. If the current site was active -- one of the four neighboring sites was randomly chosen. If the neighboring site was either empty or occupied by an A particle, the simulation step ended whereas if it was occupied by a B particle (adsorbed in an active site due to the fact that every B particle was adsorbed exclusively on the active region) an reaction attempt was realized with the probability $P_r$. If the reaction attempt succeeded, both reacting particles were removed from the lattice. Independently of the reaction attempt results, the simulation step ended.
	\end{itemize}
	\item If the lattice site chosen at the beginning of the simulation step was occupied and it contained a B particle, the simulation step ended. B particles, in contradistinction to particles A, could neither diffuse nor desorb in our model and the possibility of the surface reaction in the algorithm was taken into account when A particles were considered.
\end{itemize}

In Monte Carlo simulations some details of implementation of physical processes in the simulation algorithm are important only if quantitative results are required, e.g. if the rate constants are calculated based on the values of processes probabilities \cite{Fichthorn1991, Jansen1999}. In the presented algorithm, for example, the exclusion of the surface reaction when a B particle is found, causes that the reaction probability should be multiplied by the factor 2 if one wants to use it to calculate macroscopic quantities. In this paper we are interested only in the qualitative description of the system so absolute values of probabilities do not matter, only ratios of probabilities are important.

The first stage of each simulation was an equilibration phase. Due to the fact that at the beginning of each simulation the number and configuration of particles on the lattice in general was changing, it was necessary to wait until the system reaches a stationary state (i.e. until, for given values of pressures and probabilities, numbers of adsorbed particles of each kind and the rate of the product creation were fluctuating around constant values). After the equilibration phase an averaging phase started -- average values of numbers of adsorbed particles and the reaction rate were calculated based on a certain number of simulation steps in the stationary state.

\section{Simulations and results} \label{sec:sim_and_res}

The presented algorithm was used to simulate the dynamics of processes in the model of the catalytic process with the spillover introduced in Sec. \ref{sec:sys_desc}. Parameters of simulations were chosen to correspond to assumptions which were taken during the derivation of the analytical description of the system by Zhdanov and Kasemo in \cite{Zhdanov1997}. This means mainly that the simulations were carried out always in the high reaction rates regime and therefore the value of reaction probability was always high in comparison with probabilities of other processes. The simulations were carried out for certain sets of the model parameters to investigate the dynamics of the system for various values of reactants' partial pressures.

The results presented by Zhdanov and Kasemo \cite{Zhdanov1997} show that analytical solutions derived for extreme values of partial pressures, i.e. starting either from values of $p_A$ near 0 or near 1 and used in a moderate region of pressures lead to two different types of behavior, especially that there exists multiple steady-state and hysteretic effects may occur. To verify the hypothesis about a hysteresis and to find multiple stable stationary states, for each set of model parameters, two different series of simulations were performed. The first series was started with an empty lattice for a small, close to zero, value of $p_A$. Than, when the stable stationary-state was reached and average numbers of particles were constant, results were calculated, the value of $p_A$ was increased by a certain interval and a new simulation was started (therefore, in general, at the beginning of every next simulation the lattice was not empty). The procedure was repeated until the value of $p_A$ was near 1. The second series of simulations was carried on in an analogous way but the first simulation in these series was started for $p_A$ value near 1 and in succeeding simulations $p_A$ was decreasing. In the simulations the $p_A$ value was changing in 192 steps with a 0.05 interval, from 0.02 to 0.98 in the first type series and from 0.98 to 0.02 in the second type.

The simulations were carried on using both lattices consisting of active and inactive sites (this corresponded to a catalyst with an active metal particle on the support) and using lattices built of active sites only (this corresponded to the system where the catalytic process occurs only on an active metal particle without contribution of the support). The comparison of the results obtained from both types of simulations allows to investigate the influence of the spillover effects on the considered catalytic process. In each simulation surface coverages of an active and inactive region and the surface reaction rate were analyzed. The reaction rate was calculated as the number of product particles created in a Monte Carlo step (MC step). The MC step consisted of simulation steps described in Sec. \ref{sec:sim_alg} and the number of those steps in the MC step was equal to the number of sites in the simulation lattice.

The number of steps necessary for equilibration was determined in a series of introductory simulations and it was equal to $2\times10^4$ MC steps. However, for each first simulation in each series this number had to be increased to $10^5$ because these simulations were started with an empty simulation lattice and in this case the equilibration took more time. $1\times10^4$ MC steps after the equilibration were taken to average simulation results. Based on introductory simulations, an adsorption probability of  particles on the active region was determined as $P_{B,ads,act}=0.005$. Probability of diffusion of adsorbed A particles was taken as 1.0 independently on the kind of the starting and target lattice site (an isotropic lateral diffusion). The probability of surface reaction was equal to 0.99. Probabilities of desorption of A particles from both active and inactive sites were determined as 0.5 and 0.05 respectively. The choosing of values of the two remaining probabilities in the model (probabilities of A adsorption on active and inactive sites: $P_{A,ads,act}$ and $P_{A,ads,inact}$) allows to control the spillover process and to investigate its influence on the overall catalytic process. Adsorption and desorption probabilities in each simulation were fixed, i.e. they did not depended on the local configuration on the lattice, therefore lateral adsorbate-adsorbate interactions changing adsorption/desorption energy barriers were neglected. As mentioned in Sec. \ref{sec:sim_alg}, in a case of a qualitative analysis of simulations results, only ratios of probabilities and not their absolute values influence the dynamics of processes. Therefore, in the presented simulations, values of those probabilities were chosen to qualitatively correspond to system properties, i.e. to get the system in high reaction rates regime with the dynamics controlled by the surface diffusion. Additionally, values of simulation parameters were optimized to achieve, in sequence, the best agreement with analytical results, the smallest calculation effort and the smallest statistical errors.

\subsection{Reaction on the active metal particle} \label{subsec:react_on_particle}

In this section results of simulations with lattices consisting only of active sites are presented. The results correspond to the case of reaction occurring only on an active metal particle without influence of inactive support and therefore without the spillover effect. Probability of A adsorption was assumed as: $P_{A,ads,act} = 0.001$. The values of remaining probabilities (as described in the previous section) were: $P_{A,des,act} = 0.5$, $P_{B,ads,act} = 0.005$.

In the series with increasing value of $p_A$ surface reaction does not occur because the surface is completely poisoned by adsorbed B particles, i.e. each simulation which starts with small value of $p_A$ reaches stationary state with the surface poisoned by B particles. To investigate the mechanism of the poisoning we performed additional simulations for $p_A = 0$. In the absence of A particles ($p_A = 0$), B$_2$ adsorption is the only possible process in the system. This adsorption is irreversible, however, it cannot lead to 100\% coverage of B particles because B$_2$ particles adsorb on the surface with randomly chosen orientations and therefore some lattice sites remain unoccupied. As the adsorption of B$_2$ cannot occur on single-unoccupied sites, B coverage is less than 100\%. If surface diffusion of adsorbed B particles would be present, B coverage would be monolayer but in our model these particles were immobile. If $p_A$ is slightly higher than 0, an additional adsorption process occurs -- particles A can adsorb on single-unoccupied lattice sites. Each A adsorption act enables surface reaction which in turn can produce a pair of neighboring unoccupied sites. Therefore, because B$_2$ is still dominant in the gas phase, these unoccupied sites can be attacked by B$_2$ and the coverage of B increases. The presence of certain amount of A particles is necessary for B particles to poison the surface. Successive increasing of $p_A$ in subsequent simulations of the series cannot decrease the B coverage. Adsorption of B is irreversible and the only way to decrease the number of adsorbed B particles is the surface reaction with A, however, these particles are not only absent on the surface but also cannot be adsorbed as well due to the absence of unoccupied sites. In result, in each simulation during the series with increasing $p_A$ the surface is poisoned by B and surface reaction cannot occur.

\begin{figure}
\centering
\includegraphics[width=0.7\textwidth]{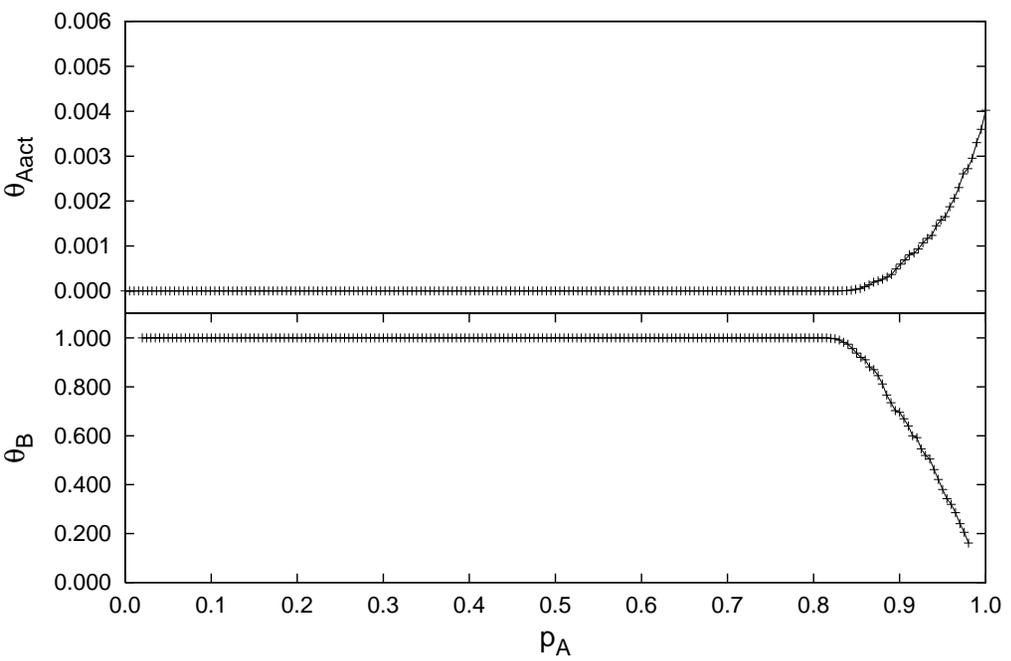}
\caption{Surface coverage of A and B particles vs. partial pressure of A in the gas phase for simulation series with the decreasing ('dec') value of $p_A$ in the case of the active metal particle without the support.} \label{fig:particle_coverage}
\end{figure}

Fig. \ref{fig:particle_rate} presents the surface reaction rate (measured as a number of product particles created in one MC step per one lattice site) vs. partial pressure of A substrate for the series with decreasing value of $p_A$. In this series reaction occurs above $p_A = 0.82$, its rate reaches maximum about $p_A = 0.93$ and instantly decreases near $p_A = 1$. In the range of pressures below 0.82 reaction does not occur. This behavior can be explained by the analysis of surface coverages. Fig. \ref{fig:particle_coverage} presents surface coverages of A and B particles vs. partial pressure of substrate A. Even though near $p_A = 1$ adsorption of A dominates, the surface is not poisoned by A. The reason of this behavior is the reversibility of adsorption. Particles A can both adsorb and desorb, therefore, the steady-state coverage is the effect of competition of these processes. For reasons explained later in this work, the choosing of adsorption and desorption probabilities ($P_{A,ads,act} = 0.001$ and $P_{A,des,act} = 0.5$) was done in such a way, to get a system with relatively small steady-state A coverages. Accordingly, the surface cannot be poisoned by A due to the adsorption reversibility and the maximum A coverage is relatively low (about 0.004) -- even at $p_A$ values near 1 -- due to the ratio of adsorption/desorption probabilities.

Adsorption of B$_2$ particles is irreversible, therefore below $p_A = 0.82$ the surface is poisoned by B. The mechanism leading to 100\% coverage is the same as in the simulations starting from $p_A$ near 0. But here the reason of effectively low A coverage just before the poisoning is the consumption of A adsorbate by surface reaction. Inside the reaction window B coverage is less than monolayer due to surface reaction and domination of A in the gas phase.

Simulations of reaction on the active metal particle can be compared with ZGB model. Both models show B poisoning region for $p_A$ values below the reaction window, however, in our case A poisoning above the reaction window does not occur due to the A adsorption reversibility. The reaction window in our model is shifted toward high values of $p_A$ with respect to ZGB because the steady-state A coverage is lower than in the ZGB model.

The presented results show that for the regions of $p_A$ above 0.82 at least two stable stationary states exist in the system: the state in which the reaction occurs (reached in the 'dec' series) and the state with the surface poisoned by B (reached in the 'inc' series). If the simulations start with an empty surface, the system would always reach the stationary state with lower value of the surface B coverage because the final stationary state depends on initial conditions of the system. In the case of experimental investigations of such systems the situation would be the same, if one started an experiment with the surface not covered by an adsorbate or covered by only one reactant (see \cite{Libuda2001, Hoffmann2001}), one would reach finally only one stable stationary state. In our simulations, starting from both low and high values of $p_A$ caused that two stable stationary states could be reached.

\subsection{Reaction on the active metal particle with the support and spillover effect} \label{subsec:react_with_spillover}

\begin{figure}
\centering
\includegraphics[width=0.7\textwidth]{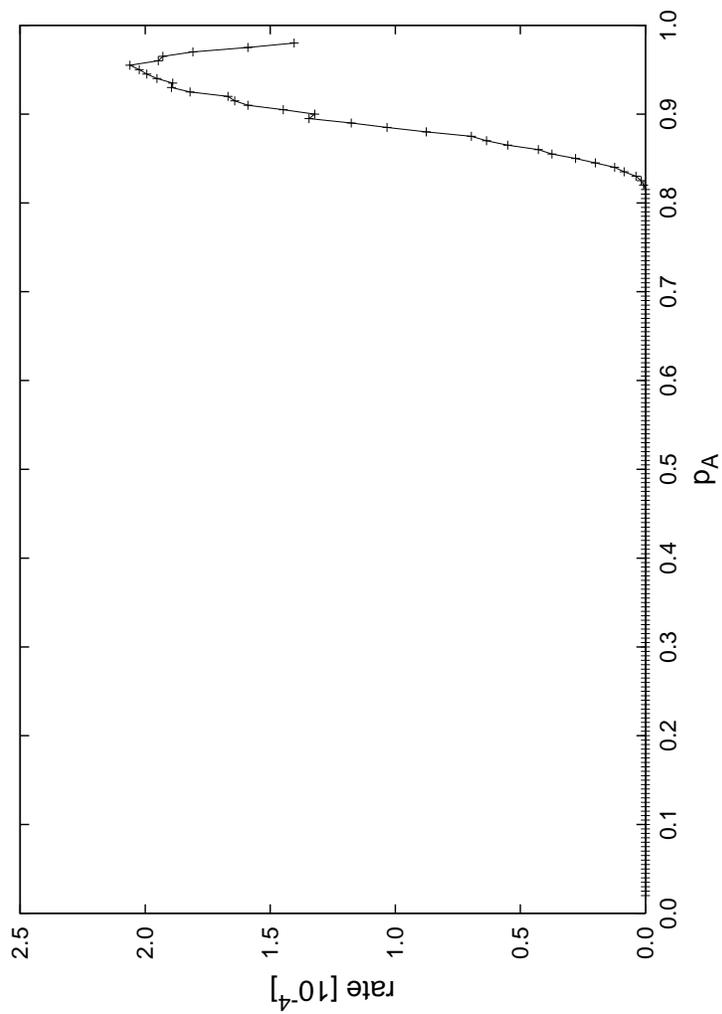}
\caption{Surface reaction rate vs. partial pressure of A in the gas phase for simulation series with the decreasing ('dec') value of $p_A$ in the case of the active metal particle without the support.} \label{fig:particle_rate}
\end{figure}

\begin{figure}
\centering
\includegraphics[width=0.7\textwidth]{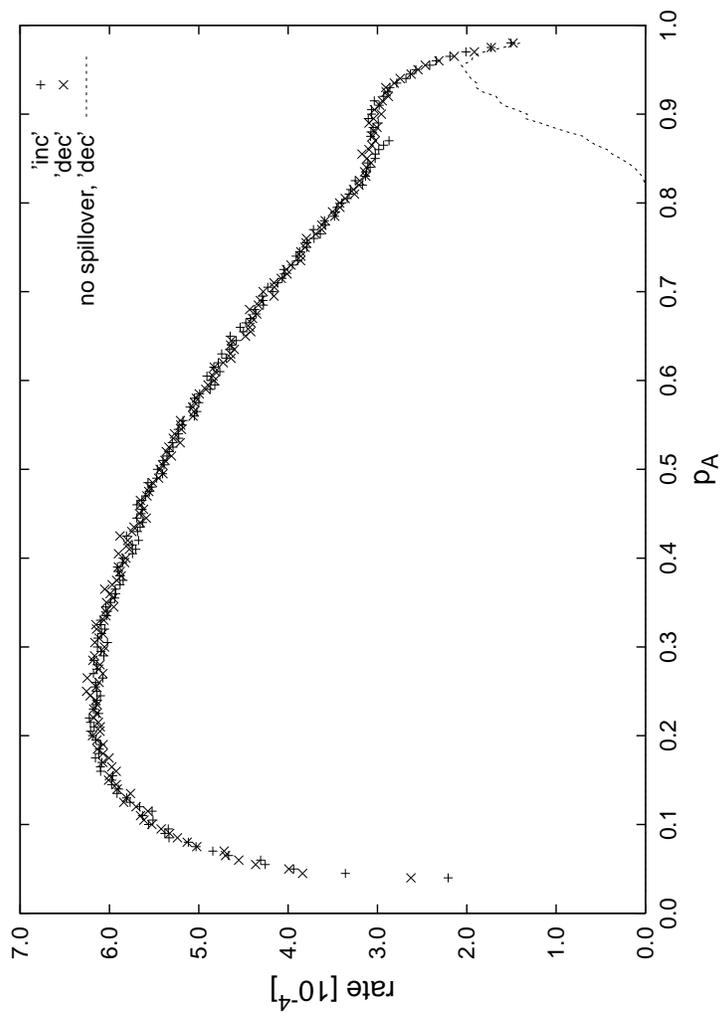}
\caption{Surface reaction rate vs. partial pressure of A in the gas phase for simulation series with the increasing ('inc') and decreasing ('dec') value of $p_A$ in the case of the active metal particle with the support and spillover effects.} \label{fig:spill_rate}
\end{figure}

In this section results of simulations with lattices consisting of both active and inactive sites are presented. Those results correspond to the case of the reaction occurring on a supported metal catalyst where a metal particle is placed on the inactive support and spillover effects occur. Probabilities of adsorption were assumed as following: $P_{A,ads,act} = 0.001$, $P_{A,ads,inact} = 0.8$, $P_{A,des,act} = 0.5$, $P_{A,des,inact} = 0.05$, $P_{B,ads,act} = 0.005$. Probabilities of processes occurring on the active region were equal to values taken in Sec. \ref{subsec:react_on_particle}. Fig. \ref{fig:spill_rate} presents the surface reaction rate (measured as a number of product particles created in one MC step per one active lattice site) vs. partial pressure of A substrate. Results of two series of simulations: with increasing (marked as 'inc') and decreasing ('dec') value of $p_A$ are shown. For comparison, results for the process with no spillover (taken from Fig. \ref{fig:particle_rate}) are also presented.

In distinction from the simulations with active lattice sites only, in presence of spillover, reaction occurs in both series of simulations (with increasing and decreasing values of $p_A$). Reaction occurs in very wide range of $p_A$ values and the reaction rate is about 3 times higher with respect to the case of active sites only. Additionally, no differences between results of each series ('inc' and 'dec') were observed.

\begin{figure}
\centering
\includegraphics[width=0.7\textwidth]{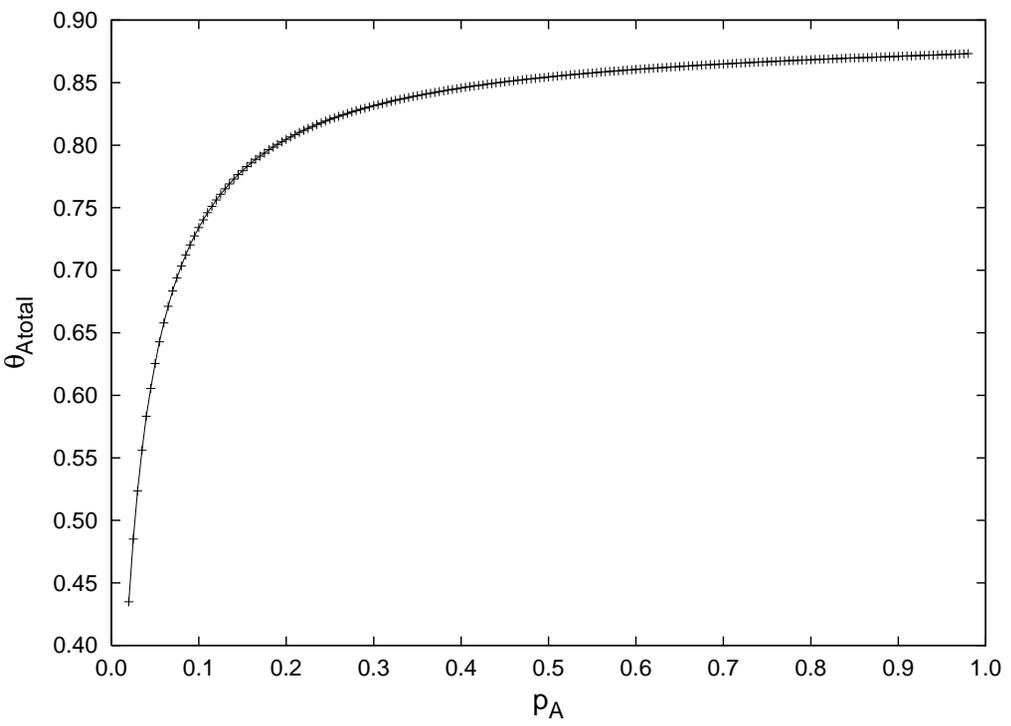}
\caption{The whole surface coverage of A particles vs. partial pressure of A in the gas phase in the case of the active metal particle with the support and spillover effects.} \label{fig:spill_natotal}
\end{figure}

\begin{figure}
\centering
\includegraphics[width=0.7\textwidth]{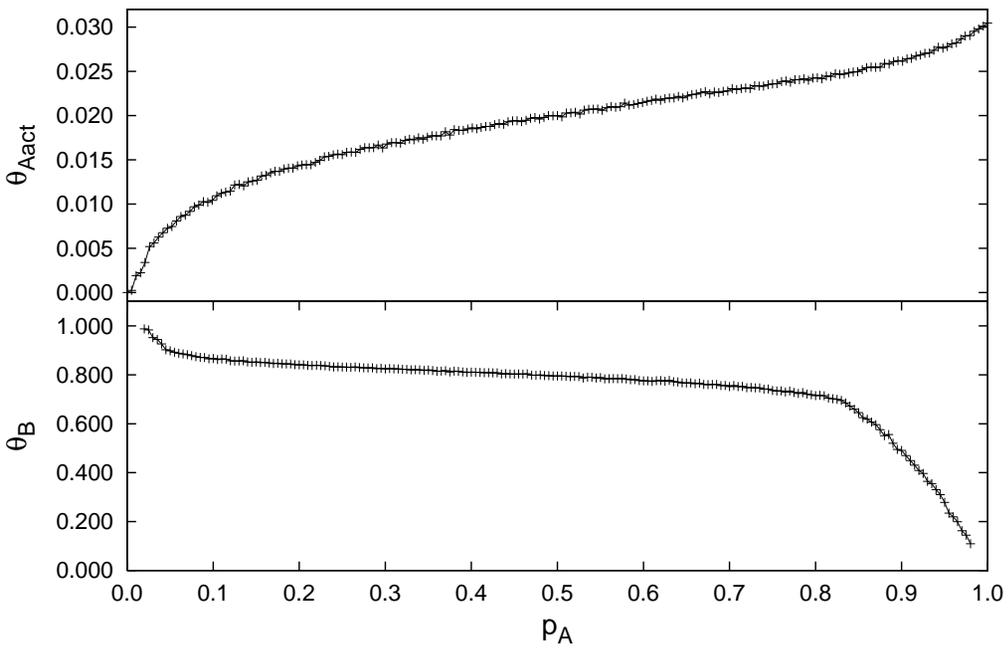}
\caption{The active region coverage of A and B particles vs. partial pressure of A in the gas phase in the case of the active metal particle with the support and spillover effects.} \label{fig:spill_particle_coverage}
\end{figure}

As in the previous case, one can find an explanation analyzing surface coverages. Fig. \ref{fig:spill_natotal} presents A coverage of the whole surface - both active and inactive region (measured as a number of adsorbed A particles per one lattice site) vs. partial pressure of A substrate. Fig. \ref{fig:spill_particle_coverage} presents the coverages of the active region by A and B particles (measured as a number of adsorbed particles per active lattice site) vs. partial pressure of A substrate.

The curve on Fig. \ref{fig:spill_natotal} is an isotherm of Langmuir type because about 89\% of lattice sites are inactive and on the inactive region only unimolecular reversible adsorption of A occurs. The maximum A covarage (for $p_A$ near 1) is lower than 100\% due to the reversibility of A adsorption but is much higher than the maximum A coverage on the active region because of the higher ratio of A adsorption/desorption probabilities ($P_{A,ads,inact} / P_{A,des,inact} > P_{A,ads,act} / P_{A,des,act}$).

In the presence of the inactive support, independently on the $p_A$ value, the active region of the surface is not poisoned by B particles. The escape of B particles to the inactive support cannot be the reason of this effect because, according to assumptions of the model, B particles do not diffuse. Therefore the only reasonable explanation of this effect is the diffusion of A particles from the inactive support to the active region and the removing of B articles by surface reaction. This explanation is also supported by the fact, that the A coverage of active region in the current model is higher with respect to the model without spillover (compare Fig. \ref{fig:spill_particle_coverage} and \ref{fig:particle_coverage}).

Even for low values of $p_A$ and high values of B coverage of the active region, the reaction occurs because the A substrate is provided from the support due to the surface diffusion and, at least at the border between the active and inactive region, the reaction always occurs. It leads to the decrease of B coverage on the active region and therefore A particles can adsorb directly on active sites from the gas phase. The effect of the lower than 100\% B coverage on supported metal catalysts was observed experimentally \cite{McCabe1977, Conrad1978}. In previous Monte Carlo studies the effect was either neglected or included by artificial assumptions in the simulation algorithm \cite{Garcia2001}. In the case of spillover only one stable stationary states of surface coverages was found.

The comparison between Fig. \ref{fig:particle_rate} and Fig. \ref{fig:spill_rate} allows to conclude that the presence of the inactive support and diffusion of A particles from and to the active region causes three changes in the system: a new high band with a maximum located about $p_A = 0.25$ occurs, the local rate maximum in the range of high values of $p_A$ increases and this maximum is shifted towards lower $p_A$ values. We conclude that the reaction in the range of high values of $p_A$ is governed by the same mechanism as in the system without the inactive support. However, for low values of $p_A$ the reaction is supported by the surface diffusion from the inactive support. In other words, in the range of high $p_A$ the reaction is governed by the adsorption channel, and for low $p_A$ -- by the spillover channel. These conclusions differ from ones obtained from an analytical description by Zhdanov and Kasemo \cite{Zhdanov1997} because in our case in the presence of inactive support a band that comes from the adsorption channel still exists for high values of $p_A$. The system with the inactive support and the spillover keeps properties of the system without support and shows those properties in the case of high values of the A substrate pressure. Similar results were obtained by Kovalyov et al. in \cite{Kovaliov2003}. Our model also does not show histeretic behavior in the region of moderate $p_A$ values.

We conclude, that the increasing of the local maximum appearing in the range of high pressures of A with the respect to the model without spillover, i.e. the increasing of the adsorption channel efficiency, is caused by the appearing of empty sites in the active region due to the additional removing of adsorbed particles by the surface reaction driven by spillover. The additional empty sites increase the rate of A adsorption on the active region and therefore strengthen the adsorption channel. The same mechanism is responsible for the shifting of the maximum toward lower $p_A$ values. The strengthening of the adsorption channel manifests itself not only at the position of the maximum but increases the rate of A adsorption in the whole range of pressures. The shifting is caused by the asymmetry of the strengthening, the reaction rate in the region governed by the adsorption channel is increased stronger for lower values of $p_A$.

\begin{figure}
\centering
\includegraphics[width=0.7\textwidth]{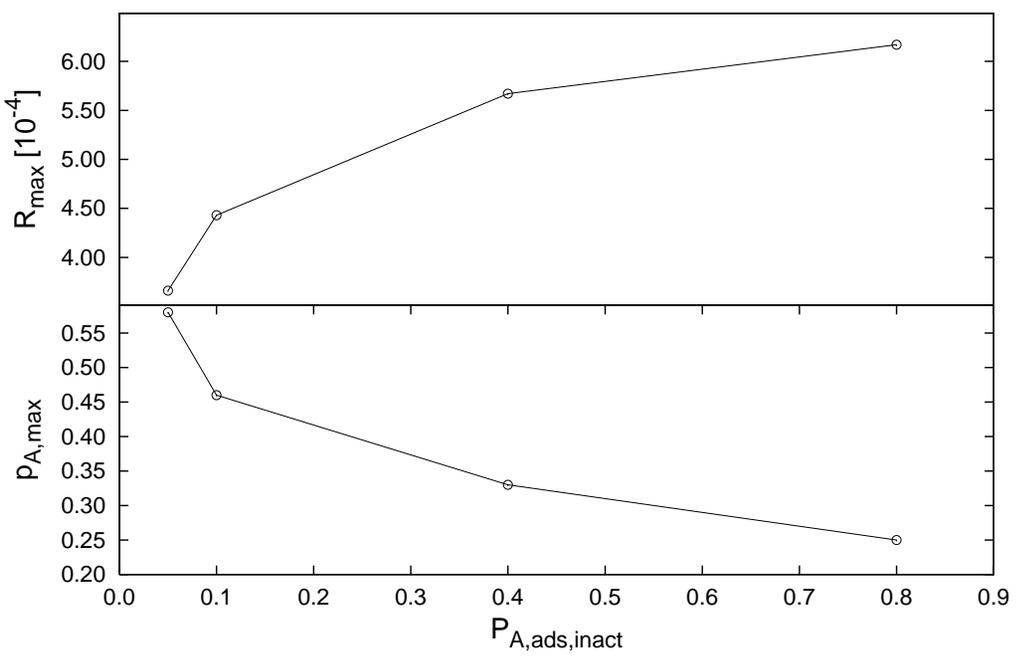}
\caption{The maximum reaction rate and the partial pressure of A reactant corresponding to the maximum reaction rate vs. probability of A adsorption on the inactive support.} \label{fig:spill_maxs}
\end{figure}

To investigate more precisely the influence of spillover channel on the behavior of both bands occurring on the rate vs. $p_A$ plot, a series of simulations analogous to previous one but for different values of A adsorption probability on the inactive region were carried on. The increasing of this probability leads to the increasing of the steady-state coverage of inactive region by A particles and increases the flux of A particles from the support onto the active region. Fig. \ref{fig:spill_maxs} presents the maximum reaction rate and the $p_A$ pressure corresponding to the maximum obtained in series of simulations for one set of model parameters (i.e. the maximum rate obtained on the plots analogous to Fig. \ref{fig:spill_rate}) vs. probability of adsorption of A particles on the support (all model parameters for simulations presented on this plot except $P_{A,ads,inact}$ were fixed). The points for $P_{A,ads,inact} = 0$ are taken from simulations of the system without support i.e. for the system without spillover effects.

\begin{figure}
\centering
\includegraphics[width=0.7\textwidth]{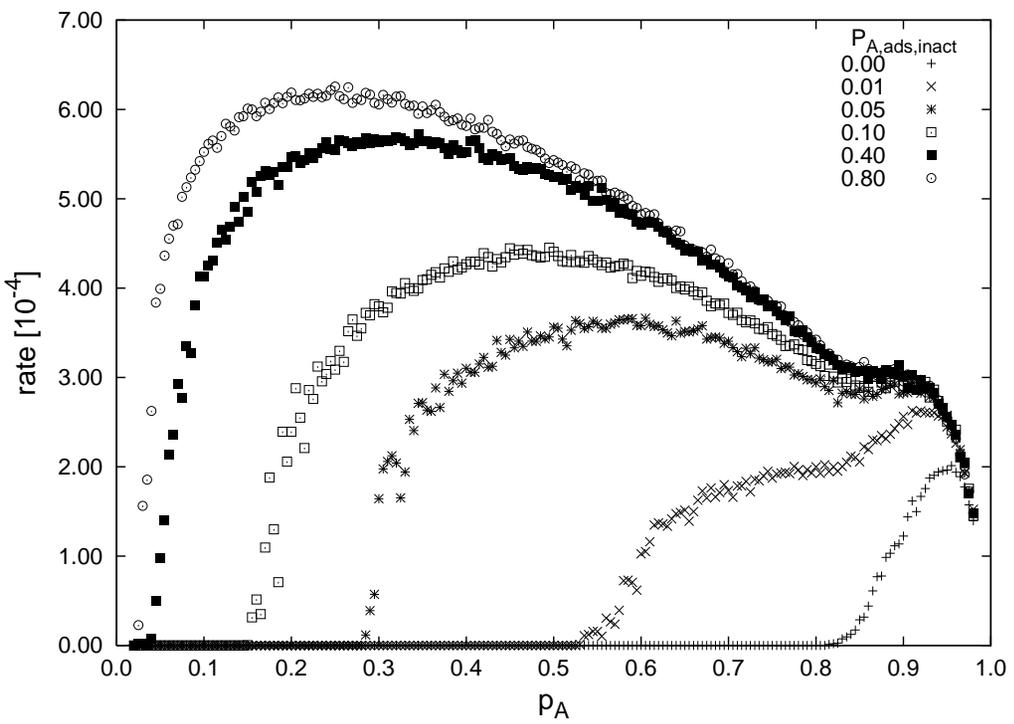}
\caption{The reaction rate vs. partial pressure of A reactant for different values of A adsorption probability on the inactive support for 'dec' simulations series.} \label{fig:spill_rshift}
\end{figure}

Fig. \ref{fig:spill_maxs} shows that the increasing influence of spillover channel leads to the increasing of the maximum reaction rate whereas the $p_A$ pressure corresponding to that maximum, is decreasing. When the spillover becomes more important, the reaction rate increases in the range of low $p_A$ pressures. In the case of weak spillover, the system acts as a system consisting of active centers without a support. This effect is shown on Fig. \ref{fig:spill_rshift} which presents the reaction rate vs. $p_A$ for different values of probability of A particles adsorption on the support.

\subsection{Spillover effects with the strong adsorption channel} \label{subsect:strong_ads_channel}

\begin{figure}
\centering
\includegraphics[width=0.7\textwidth]{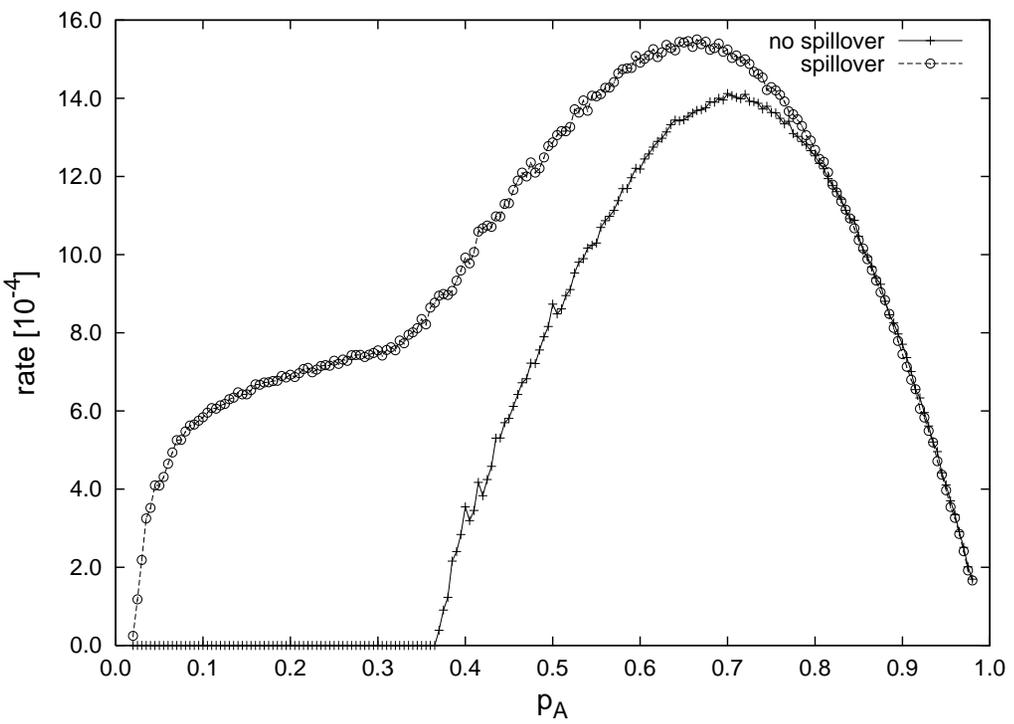}
\caption{The surface reaction rate vs. A partial pressure in the gas phase in the case of strong adsorption channel.} \label{fig:strong_ads_r}
\end{figure}

Among various models considered in the present work (with different values of all parameters) it is worth to present the case where an adsorption of A substrate on the active region is much faster than in simulations presented in the previous section. Fig. \ref{fig:strong_ads_r} presents the reaction rate vs. pressure of A substrate for the model with the probabilities: $P_{A,ads,act} = 0.01$, $P_{A,ads,inact} = 0.8$, $P_{A,des,act} = 0.5$, $P_{A,des,inact} = 0.05$, $P_{B,ads,act} = 0.005$. Here, $P_{A,ads,act}$ probability is 10 times larger than in the simulations presented in previous sections. Reaction rate for both the system consisting of active sites only (i.e. without spillover effects) and for system consisting of the active region and the inactive support (with spillover) are presented.

In this case, when fast adsorption on inactive region is present, the reaction rate is high even in the system consisting only of active sites whit respect to simulations presented in Sec. \ref{subsec:react_on_particle}. The range of $p_A$ values for which the reaction occurs is much wider than in the cases presented above and the rate maximum is shifted toward lower $p_A$ values (compare with Fig. \ref{fig:particle_rate}). The presence of the inactive support slightly increases the maximum reaction rate and shifts the maximum left. Additionally, spillover broadens the range of active pressures and a new band occurs below $p_A = 0.3$. But since reaction on the isolated active region is very fast, the height of the band located above $p_A = 0.3$ is still bigger than the height of the band coming from the spillover channel. This effect is different than in the cases presented in previous sections because now, in contradiction to Fig. \ref{fig:spill_rshift}, even for relatively low values of the ratio $P_{A,ads,inact} / P_{A,ads,act}$ (this ratio corresponds to the strength of spillover channel in comparison with the adsorption channel and on Fig. \ref{fig:strong_ads_r} is equal 80) the reaction rate in the range of low A pressures increases and the reaction occurs always in the whole range of $p_A$ values. In previous simulations the reaction occurred for the whole range of $p_A$ values only if $P_{A,ads,inact} \geq 0.4$ (then $P_{A,ads,inact} / P_{A,ads,act} \geq 400$), i.e. for very strong spillover channel.

Results for our model with strong adsorption channel may be compared with results presented by Kovalyov et al. in \cite{Kovaliov2003}. In that work the presence of surface diffusion from and to the inactive support causes slight increasing of the maximum reaction rate and a new, but not well resolved, band occurs for low $p_A$ values. Those results are comparable with ours only qualitatively due to the differences in the values of probabilities chosen in both models.

It is noteworthy that for certain values of A adsorption probabilities on both the active and inactive surface region one can get the system in which the reaction rate will be almost independent of reactants' partial pressures. Such an experimental system with CO oxidation on Pt/SnO$_2$ catalyst was reported, for example, by Grass and Linz \cite{Grass1997} but in that work the system behavior was explained using a different model (a reaction mechanism including an oxygen migration was proposed).

\section{Conclusions} \label{sec:conclusions}

Supported metal catalysts show both energetic and spatial heterogeneities. Dynamic Monte Carlo simulations can treat those inhomogeneities explicitly. In this article results of Dynamic Monte Carlo simulations of the model reaction of CO oxidation on the supported metal catalyst are presented. In this reaction the influence of oxide support is significant because of the presence of spillover effects and therefore the importance of spillover was investigated. We simulated catalytic systems both with and without spillover effects and in the former case the importance of spillover was controlled by adjusting the spillover contribution.

\begin{enumerate}

\item In systems without spillover:

\begin{enumerate}
	\item The poisoning by A particles does not occur due to the A adsorption reversibility.
	\item The catalyst surface cannot be poisoned by the B reactant if reactant A is not present because of the two-center dissociative adsorption of B. Only if reactant A is present and its partial pressure is low, B particles can poison the catalyst. \label{concl:part_1}
	\item Two stable stationary states of B coverage were found \label{concl:part_2}
	\item The reaction rate achieves maximum for relatively high values of $p_A$. \label{concl:part_3}
\end{enumerate}

\item In systems with spillover:

\begin{enumerate}
	\item The surface cannot be poisoned by B particles for any value of $p_A$. If $p_A = 0$ the effect is caused by two-center adsorption whereas for $p_A > 0$ the reason is the constant presence of A particles on the active region due to diffusion from the support. We have obtained this effect without additional assumptions in the algorithm as in the previous studies. \label{concl:spill_1}
	\item The reaction is possible for every value of $p_A$ (except 0 and 1) because the surface cannot be poisoned. \label{concl:spill_2}
	\item Depending on the value of $p_A$, reaction occurs in two regimes: driven by the spillover channel and driven by the direct adsorption of A on the active region of the surface. \label{concl:spill_3}
	\item No hysteretic behavior have been found. \label{concl:spill_4}
	\item Relative values of reaction rate in the case of reaction  driven by the spillover channel and reaction driven by the direct adsorption of A on the active region depend on relative values of probabilities of adsorption and desorption of A on the support and on the active region of the catalyst surface. For certain values of those probabilities one can get a system in which the reaction rate will be almost independent of partial pressures of reactants. \label{concl:spill_5}
\end{enumerate}

\end{enumerate}

Some of our results differ qualitatively from the ones obtained by Zhdanov and Kasemo in \cite{Zhdanov1997} for an analytical model: we have not found hysteresis in the system with spillover and we have found two reaction regimes i.e. driven by spillover and driven by adsorption because we were able to use one model for the whole range of values of reactants' partial pressures. Analytical solutions require additional assumptions about partial pressures and surface geometry, whereas Dynamic Monte Carlo simulations can be used for wide variety of model parameters and can treat surface heterogeneity explicitly.

Our results are in good qualitative agreement with the Kovalyov and co-workers' results \cite{Kovaliov2003}. Results for their model with spillover in the case of flat and uniform active particle are comparable with our simulations for strong adsorption channel. However, our data for a strong spillover channel are qualitatively new.

Dynamic Monte Carlo simulations are a useful tool to study processes occurring on heterogeneous surfaces, but because of the lack of experimental data which are needed to calculate model parameters, in most cases their results should be treated only qualitatively. Therefore results presented in this work show qualitatively what kind of effects can occur in systems with spillover and which parameters influence them.

\clearpage

\clearpage

\begin{thebibliography}{00}



\bibitem{Conner1995}
W. C. Conner, J. L. Falconer, Chem. Rev. 95 (1995) 759.


\bibitem{Broadbelt2000}
L. J. Broadbelt, R.Q. Snurr, Appl. Catal. A 200 (2000) 23.

\bibitem{Zhdanov2000}
V. P. Zhdanov, B. Kasemo, Surf. Sci. Rep. 39 (2000) 25.

\bibitem{Zhdanov1997}
V. P. Zhdanov, B. Kasemo, J. Catal. 170 (1997) 377.

\bibitem{Libuda2001}
J. Libuda, I. Meusel, J. Hoffmann, J. Hartmann, L. Piccolo, C. R. Henry, H.-J. Freund, J. Chem. Phys. 114 (2001) 4669.

\bibitem{Hoffmann2001}
J. Hoffmann, I. Meusel, J. Hartmann, J. Libuda, H.-J. Freund, J. Catal. 204 (2001) 378.

\bibitem{Kovaliov2003}
E. V. Kovaliow, E. D. Resnyanskii, V. I. Elokhin, B. S. Bal'zhinimaev, A. V. Myshlyavtsev, Phys. Chem. Chem. Phys. 5 (2003) 784.

\bibitem{Ziff1986}
R. M. Ziff, E. Gulari, Y. Barshad, Phys. Rev. Lett. 56 (1986) 2553.

\bibitem{Fichthorn1991}
K. A. Fichthorn, W. H. Weinberg, J. Chem. Phys. 95 (1991), 1090.

\bibitem{Jansen1999}
A. P. J. Jansen, J. J. Lukkien, Cat. Today, 53 (1999) 259.


\bibitem{Grass1997}
K. Grass, H.-G. Lintz, J. Catal. 172 (1997) 446.

\bibitem{McCabe1977}
R. W. McCabe, L. D. Schmidt, Surf. Sci. 66 (1977) 101.

\bibitem{Conrad1978}
H. Conard, G. Ertl, J. Kuppers, Surf. Sci. 76 (1978) 323.

\bibitem{Garcia2001}
F. Garcia, E. E. Wolf, Chem. Eng. Journal 82 (2001) 291.

\end{thebibliography}
\end{document}